\documentclass[aps,prl,letterpaper,11pt,twoside,tightenlines,nofootinbib,showpacs,preprint,twocolumn]{revtex4}
\usepackage{graphicx}
\usepackage[sort&compress]{natbib}
\usepackage{latexsym}
\usepackage{epsfig}

\newcommand{\be}{\begin{equation}}

\newcommand{\ee}{\end{equation}}
\newcommand{\bea}{\begin{eqnarray}}
\newcommand{\eea}{\end{eqnarray}}
\newcommand{\bef}{\begin{figure}}
\newcommand{\eef}{\end{figure}}
\newcommand{\bce}{\begin{center}}
\newcommand{\ece}{\end{center}}
\def\lsim{\mathrel{\rlap{\lower4pt\hbox{\hskip1pt$\sim$}}
    \raise1pt\hbox{$<$}}}         %less than or approx. symbol
\def\gsim{\mathrel{\rlap{\lower4pt\hbox{\hskip1pt$\sim$}}
    \raise1pt\hbox{$>$}}}         %greater than or approx. symbol
\begin{document}
\title{Symmetry Restoration by Acceleration}
\author{P. Castorina and M. Finocchiaro}
\affiliation{Dipartimento di Fisica, Universit\`a di Catania,
 and INFN Sezione di Catania, Via Santa Sofia 64
I-95123 Catania, Italia}

\date{\today}
\begin{abstract}
The restoration of spontaneous symmetry breaking for a scalar field theory for an accelerated observer is discussed by the one-loop effective potential calculation and by considering the effective potential for composite operators. Above a critical acceleration, corresponding to the critical restoration temperature,$T_c$, for a Minkowski observer by Unruh relation , i.e. $a_c/2\pi=T_c$, the symmetry is restored. This result confirms other recent calculations in effective field theories that symmetry restoration can occur for an observer with an acceleration larger than some critical value.
From the physical point of view, a constant acceleration is locally equivalent to  a gravitational field and the critical acceleration to restore the spontaneous symmetry breaking corresponds to a huge gravitational effect which, therefore,  prevents boson condensation.

\end{abstract}
 \maketitle

\section{Introduction}

In quantum field theory in flat space-time a spontaneously broken symmetry can be restored above some critical temperature \cite{rest1,rest2,rest3}.

On the other hand, whenever the background geometry is endowed with a black-hole or an event horizon, the related vacuum physically behaves like a thermal bath of quanta with a temperature , $T_H$, proportional to the surface gravity \cite{hawking}.

Therefore one expects that symmetry restoration occurs also in strong enough gravitational fields, i.e. near a horizon \cite{spallucci1}.
On the other hand, the near horizon approximation of a black-hole metric corresponds to a Rindler metric, i.e. the metric of  an observer with constant acceleration , $a$, equal to the surface gravity and Unruh \cite{unruh,crispino} showed that for any accelerated observer there is an intrinsic thermal nature of the ground state :
he/she feels a thermal bath with temperature $T=a/2\pi$.

Moreover, a broken symmetry can be also restored if a classical external field strength ( a magnetic field, for example) exceeds a critical value. Infact a particle coupled with an external field suffers a proper acceleration depending on the field strength.

The pervious discussion clearly indicates that a restoration of the symmetry can occur for an observer with an acceleration larger than some critical value
independently on the specific dynamical mechanism that produces the acceleration.

Indeed in ref. \cite{giapponesi} it has been shown that for the Nambu - Jona Lasinio (NJL) model in an accelerated frame, the chiral symmetry (broken for
$a=0$) is restored for acceleration larger than a critical value $a_c$, corresponding to $a_c/2\pi=T_c$, where $T_c$ is the critical temperature for the restoration of the symmetry due to standard ( flat space-time) thermal fluctuactions. 

In ref. \cite{zuk} the behavior of quark and diquark condensates  as seen by an accelerated observer has been studied and critical values of the acceleration for the restoration of chiral and color symmetries have been estimated.

The dissociation of mesons, described as rotating string in Rindler space,  by acceleration has been analyzed in ref. \cite{kasper} with the conclusion that
above a critical acceleration $a_c \simeq \sqrt{\sigma/J}$ , where $\sigma$ is the string tension and $J$ is the angular momentum, mesons undergo dissociation.

In this letter we discuss the restoration of spontaneous symmetry breaking for $\lambda \phi^4$ theory with similar results: above a critical acceleration, corresponding to the critical restoration temperature by Unruh relation , i.e. $a_c/2\pi=T_c$, the symmetry is restored. The calculations are based on the 
one-loop effective potential evaluation and its generalization for composite operators \cite{cjt} (CJT)  which gives the resummation of an infinite set of diagrams by self-consistent gap equations.

Initially (Sec.1) we  recall briefly some general features of a scalar field theory in Rindler metric. In Sec.2  we discuss  the one-loop calculation of the effective potential for  $\lambda \phi^4$ theory for an accelerated frame  and the results by the CJT method. General considerations about the symmetry breaking due to acceleration and the Hawking-Unruh radiation are in the final section devoted to comments and conclusions. 

\section{\Large{\bf 1. $\lambda \phi^4$} in Rindler metric}

The action for the $ \lambda\phi^{4} $ scalar theory in Rindler spacetimes \cite{crispino} can be written as follow:
\begin{equation}
I(\phi) = \int d^{4}x\sqrt{-g} \frac{1}{2}\partial_{\mu}\phi g^{\mu\nu}\partial_{\nu}\phi - \frac{1}{2}m^{2}\phi^{2} - \frac{\lambda}{4!}\phi^{4}
\end{equation}
where  $x=(\eta,\rho,\mathbf{x}_{\bot})$, $\mathbf{x}_{\bot}\equiv\,(x , y)$ and  the Rindler metric tensor $ g_{\mu\nu} $ is given by :
\begin{equation}
ds^{2} = \rho^{2}d\eta^{2} - d\rho^{2} - d\mathbf{x}^{2}_{\bot}  
\end{equation}
\begin{equation}
g_{\mu\nu} = \textmd{diag}(\rho^{2} , -1 , -1 , -1) \quad g = \textmd{Det} g_{\mu\nu}  
\end{equation}

Since the two Rindler wedge are causally disconnected from each other \cite{crispino} 
we restrict our consideration to the right Rindler wedge. Calculation in the left  wedge can be performed in the same way.

The Klein-Gordon equation for the scalar field $ \phi(x) $ is
\begin{equation}
(\partial_{\mu}\sqrt{-g}\partial^{\mu} + \sqrt{-g} m^{2})\phi(x) = 0
\end{equation}

Taking the Fourier transform with respect to $ (\eta , \mathbf{x}_{\bot}) $:
\begin{equation}
\phi(x) = \int\frac{d\omega\,d^{2}\mathbf{k}_{\bot}}{(2\pi)^{3}}e^{i(\mathbf{k}_{\bot}\cdot\mathbf{x}_{\bot} - \omega\eta)}\phi(\omega , \rho),
\end{equation}
the normalized solution turns out
\begin{equation}
\phi(\omega , \rho) = 2\sqrt{\frac{\omega\sinh(\pi\omega)}{\pi}}K_{i\omega}(\hat \alpha \rho)
\end{equation}
where
$K_{i\omega}(\hat \alpha\rho)$ is the modified Bessel function of second kind and $\hat \alpha = \sqrt{k_{\bot}^{2} + m^{2}} $.

The two-point Green's function of the free scalar field in the right Rindler wedge, $\mathcal{G}_m(x, x',q^{2})$ , is defined by the equation:
\small{\begin{equation}
[\partial_{\mu}\sqrt{-g}g^{\mu\nu}\partial_{\nu} + \sqrt{-g} m^{2}]\mathcal{G}_m(x,x',q^{2}) = \frac{1}{\sqrt{-g}}\delta(x - x')
\label{greenfunction}
\end{equation}}

The Fourier trasform of $\mathcal{G}_m(x,x',q^{2})$ with respect to $ (\eta , \mathbf{x}_{\bot}) $ is given by
$$
\mathcal{G}_m(x , x' , q^{2}) = \mathcal{G}_m(\eta - \eta' , \rho , \rho' , \mathbf{x}_{\bot} - \mathbf{x'}_{\bot} , q^{2})=
$$

\small{\begin{equation}
= \int\frac{d\omega}{2\pi}\int\frac{d^{2}\mathbf{k}_{\bot}}{(2\pi)^{2}} f_{\omega,\vec k_{\bot}} (\vec x_\bot, \vec x^{'}_{\bot},\eta,\eta^{'})\mathcal{G}_m(\omega , \rho , \rho' , \mathbf{k}_{\bot} , q^{2})
\end{equation}}
where
\begin{equation}
f_{\omega,\vec k_{\bot}} (\vec x_{\bot}, \vec x^{'}{_\bot},\eta,\eta^{'})= e^{i[\mathbf{k}_{\bot}\cdot(\mathbf{x}_{\bot} - \mathbf{x'}_{\bot}) - \omega(\eta - \eta')]} \nonumber \\
\end{equation}
By substituting back into Eq. (\ref{greenfunction}) one finds:
\begin{equation}
\Delta_R^m \mathcal{G}_m(\omega , \rho , \rho' , \mathbf{k}_{\bot} , q^{2}) = \frac{1}{\rho}\delta(\rho - \rho')
\end{equation}
where 
\begin{equation}
\Delta_R^m = [-\frac{\omega^{2}}{\rho^{2}} - \frac{1}{\rho}\frac{\partial}{\partial\rho} - \frac{\partial^{2}}{\partial\rho^{2}} + (k^{2}_{\bot} + m^{2})]
\end{equation}

The world line in Rindler coordinates of a uniformly accelerated observer with proper constant acceleration $ a $ is given as,
\begin{equation}
\eta = a\tau\qquad \rho = \frac{1}{a}\qquad \mathbf{x}_{\bot} = const
\end{equation}
and it has been generally proved \cite{uw} that Euclidean two point functions in Rindler coordinates are periodic in the direction of time with period $ a $. 

Since the Euclidean Rindler spacetime has a singularity at $ \rho = 0  $ one requires that the period of the imaginary time is $ \beta = 2\pi/a $. With this particular choice the Euclidean formalism in Rindler coordinates coincides with the finite temperature Matsubara formalism and therefore the following substitutions will be necessary in evaluating the effective potential in the next sections: 

\begin{equation}
\int\frac{dk_{0}}{2\pi} = \int\frac{d\omega}{2\pi}\,\rightarrow\, \sum_{n}\frac{1}{\beta}\quad 
\end{equation} 
\begin{equation}
k_{0} = \omega\,\rightarrow\, i\omega_{n}\quad \omega_{n}  = \frac{2n\pi}{\beta} = n \nonumber 
\end{equation}
\begin{equation}
\mathcal{G}_m(\omega,\rho,\rho', \mathbf{k}_{\bot} , q^{2})\,\rightarrow\, -\mathcal{G}_m(\omega_{n} , \rho , \rho' , \mathbf{k}_{\bot}  , q^{2}) 
\end{equation}

\section{2. Effective potential for accelerated observer}

Let us now study the effective potential for a Rindler observer when for a Minkowski observer the effective potential is assumed to possess a symmetry breaking.
For classical constant field configuration, $\phi_c$, the one loop effective action in Rindler coordinates turns out to be:
\begin{equation}
\Gamma[\phi_{cl}] =  I(\phi_{cl}) + \frac{1}{2}i\hbar\ln Det(\partial_{\mu}\sqrt{-g}\partial^{\mu} + \sqrt{-g}M^{2})
\end{equation}
$$
= I(\phi_{cl}) +  \frac{1}{2}i\hbar\int d^{4}x\ln(\partial_{\mu}\sqrt{-g}\partial^{\mu} + M^{2}\sqrt{-g})
$$ 
where $M^2 = m^2 + \lambda  \phi_{cl}^2$ and $I(\phi_{cl})$ is given by 
\begin{equation}
 I(\phi_{cl}) = -\int d^{4}x\sqrt{-g}(\frac{1}{2}m^{2}\phi_{cl}^{2} + \frac{\lambda}{4!}\phi_{cl}^{4} ). 
\end{equation}
The effective potential ( $\phi_{cl} = \phi_0$, $\hbar = 1$) is defined as
\begin{equation}
V^{R}(\phi_{0}) = - \frac{\Gamma(\phi_{0})}{\int d^{4}x\sqrt{-g}} 
\end{equation}
and  by the following relation,
\begin{equation}
\int d^{4}x\ln(\partial_{\mu}\sqrt{-g}\partial^{\mu} + M^{2}\sqrt{-g})
\end{equation}
\begin{equation}
 = \int d^{4}x\sqrt{-g}\int_{0}^{M^{2}}dq^{2}\mathcal{G}_M(x , x' , q^{2}) + const,
\label{trick}
\end{equation}
where $ \mathcal{G}_M(x , x' , q^{2}) $ is defined by  previous eqs.(8-11) with the substitution $m^2 \rightarrow M^2$, 
it turns out to be
\begin{equation}
V^{R}(\phi_{0}) = \frac{1}{2}m^{2}\phi_{0}^{2} + \frac{\lambda}{4!}\phi_{0}^{4} - \frac{i}{2}\int_{0}^{M^{2}}dq^{2}\mathcal{G}_M(x , x' , q^{2}).
\label{effpot}
\end{equation}
In Euclidean space with the periodic boundary condition in eqs.(13,14) and by introducing the Fourier transform of the Green's function, eq.(8), the previous equation gives 
\begin{equation}
V^{R}(\phi_{0}) = \frac{1}{2}m^{2}\phi_{0}^{2} + \frac{\lambda}{4!}\phi_{0}^{4} 
\end{equation}
$$
+ \frac{1}{4\pi}\sum_{n} \int_{qk} f_{\omega,\vec k_{\bot}} (\vec x_{\bot}, \vec x^{'}{_\bot},\eta,\eta^{'}) \mathcal{G}_M(\rho , \rho' , q^{2})
$$
where 
\begin{equation}
\int_{qk} \equiv \int_{0}^{M^{2}}dq^{2}\int\frac{d^{2}\mathbf{k}_{\bot}}{(2\pi)^{2}}
\end{equation}
and $\mathcal{G}_M(\rho, \rho', q^{2}) \equiv -\mathcal{G}_M(\omega_{n},\rho,\rho', \mathbf{k}_{\bot}  , q^{2})$ is now solution of the equation
\begin{equation}
[\frac{\omega_{n}^{2}}{\rho^{2}} - \frac{1}{\rho}\frac{\partial}{\partial\rho} - \frac{\partial^{2}}{\partial\rho^{2}} + (k^{2}_{\bot} + M^{2})]\mathcal{G}_M(\rho , \rho' , q^{2}) = \frac{1}{\rho}\delta(\rho - \rho')
\label{egreenfunction}
\end{equation}
which , according to eq.(6) can be written as
\begin{equation}
\mathcal{G}_M(\rho , \rho' , q^{2}) = \int_{0}^{+\infty}\frac{d\omega}{2\pi}\,\frac{\phi(\omega , \rho)\,\phi(\omega , \rho')}{\omega_{n}^{2} + \omega^{2}}
\label{egfundef}
\end{equation}
Since the world line in Rindler coordinates of a uniformly accelerated observer with proper constant acceleration $ a $ is given by eq.(12),
one sets $ \eta = \eta' = a\tau $ and $ \rho = \rho' = \frac{1}{a} $ and by changing the integration variable in Eq.(\ref{egfundef}) from $ \omega $ to $ \frac{\omega}{a} $, it turns out
\begin{equation}
V^{R}(\phi_{0}) = \frac{1}{2}m^{2}\phi_{0}^{2} + \frac{\lambda}{4!}\phi_{0}^{4}
\end{equation}
$$
+ \frac{1}{2\pi\,a}\sum_{n}\int_{qk} \int_{0}^{+\infty}d\omega\,\frac{\omega}{a}\sinh\frac{\pi\omega}{a}\frac{K^{2}_{\frac{i\omega}{a}}(\frac{\alpha}{a})}{\frac{\omega^{2}}{a^{2}} + \omega_{n}^{2}}
$$
By performing the sum on the Matsubara frequency the final result for the effective potential is
\begin{equation}
V^{R}(\phi_{0}) = \frac{1}{2}m^{2}\phi_{0}^{2} + \frac{\lambda}{4!}\phi_{0}^{4} 
\end{equation}
$$
+\frac{1}{2a}\int_{qk} \int_{0}^{+\infty}d\omega\ g(\omega) K^{2}_{\frac{i\omega}{a}}(\frac{\alpha}{a})  
$$
with $ \alpha^{2} = k^{2}_{\bot} + q^{2} $  and 
\begin{equation}
g(\omega) = \cosh\frac{\pi\omega}{a}
\end{equation}

In order to calculate the critical acceleration $ a_{c} $ for symmetry restoration we will impose the following condition \cite{rest3}:
\begin{equation}
\frac{\partial\,V^{R}(\phi_{0})}{\partial\phi_{0}^{2}}\bigg\arrowvert_{\phi_{0} = 0} = 0
\label{condition}
\end{equation}
which gives
\begin{equation}
\frac{m^{2}}{2} + \frac{\lambda}{8\pi^3\,a}\int_{0}^{+\infty}d\omega\ g(\omega)\int_{0}^{+\infty}dk\,k\,K^{2}_{\frac{i\omega}{a}}(\frac{\sqrt{k^{2} + m^{2}}}{a}) = 0   
\label{gapeq}
\end{equation}
where we set $ |\vec k_{\bot}| = k $ for notation convenience.

The computation of the integrals is straithforward and one gets
\begin{equation}
\frac{m^{2}}{2} + \frac{\lambda}{16\pi^2}\int_{0}^{+\infty}d\omega\,\omega(1 + \frac{2}{e^{\frac{2\pi\omega}{a}} - 1}) = 0
\label{gapeqI}
\end{equation}
in complete analogy with the finite temperature case \cite{rest3}. Indeed by defining the renormalized mass to cancel the quadratic divergence
\begin{equation}
 m^{2} \rightarrow m^2 + \delta m^2 
\label{rmass}
\end{equation}
the critical acceleration is obtained by the equation
\begin{equation}
\frac{m^{2}}{2} + \frac{\lambda}{16\pi^2}\int_{0}^{+\infty}d\omega\,\frac{2\omega}{e^{\frac{2\pi\omega}{a}} - 1} = 0
\label{rengapeq}
\end{equation}
which, for large acceleration, gives
\begin{equation}
\frac{m^{2}}{2} + \frac{a^{2}\lambda}{32\pi^{4}}\frac{\pi^{2}}{6} = 0 
\end{equation}
i.e.
\begin{equation}
T_{c}^{2} = \frac{a_{c}^{2}}{4\pi^{2}} = -\frac{24m^{2}}{\lambda}  
\end{equation}
in agreement with  the one-loop calculation in Minkowski space-time at finite temperature \cite{rest3} ($-m^2 >0$). 

Although the final result is as expected , the calculation in not enterely trivial. Morover the determination of the critical acceleration from one loop
effective potential suffers the same infrared problem of the finite temperature calculation related with the mode $n=0$.

 As well known, in finite temperature field theory a reliable evaluation of the critical temperature requires the resummation of an infinite subset of diagrams \cite{rest3}. This can be more easily done by considering the effective potential for composite operators (CJT) \cite{cjt}, extensively applied at finite temperature, since the relevant, infinite, subset of diagrams is automatically resummed by the gap equations corresponding to the minimum conditions of the effective potential with respect to the relevan  physical  parameters of the theory.

In the analysis of the spontaneous symmetry breaking and its restoration  for $\lambda \phi^4$ theory by CJT method in the Hartree-Fock approximation ( i.e. by considering the lowest order contribution to the gap equation),  the relevant operators are $\hat \phi(x)$ and $\hat \phi(x) \hat \phi(y)$ and the corresponding parameters are the vacuum expectation value of the field and the mass in the two-point function. The calculations at finite temperature have been carried out in ref.\cite{ccz}. 

Since the gap equations in the Hartree-Fock approximation correspond to a one-loop, self consistent, calculation of the self-energy ( see \cite{ccz} for details), from our previous, explicit, one-loop calculation and from the  complete analogy of the Green's functions  between a Minkowsky observer at finite temperature  and an accelerated  observer with $T=a/2\pi$ \cite{uw}, it follows that  a more reliable evaluation of the critical acceleration $a_c/2\pi=T_c$ with respect to the one-loop result in eq.(35)  can be obtained by following the same analysis of ref. \cite{ccz}.

However the most interesting aspect is not the exact value of the critical acceleration but the restoration of the symmetry for an accelerated observer ( see \cite{uw}, sec. IV, for a different point of view).

\section{3. Comments and Conclusions}

The restoration of chiral and color symmetries in the Nambu - Jona Lasinio model for an observer with a costant acceleration above a critical value
\cite{giapponesi,zuk} and the calculation performed in the previous section clearly indicate that one can restore broken symmetries by acceleration.

Although the technical aspects of the previous calculations are sound, the physical mechanism of the restoration is unclear if one does not  recall that
a constant acceleration is locally equivalent to a gravitational field. The critical acceleration to restore the spontaneous symmetry breaking corresponds to a huge gravitational effect which  prevents boson condensation \cite{spallucci1} as in the case of a non relativistic, ideal Bose gas  \cite{cavalcanti}.

On the other hand, for accelerations  due to the observed weak gravitational fields the corresponding  Hawking-Unruh temperatures are too small to produce measurable effects. There are very interesting attempts to find gravity-analogue of the Hawking-Unruh radiation \cite{liberati,iorio} and, in our opinion, high energy particle physics seems the more promising sector to observe this effect. Indeed,  a temperatute $T \simeq 170$ Mev , corresponding to an acceleration $O(10^{35}$ $cm/s^2)$, has be reached in relativistic heavy ion collisions and the hadronic production can be understood as Hawking-Unruh radiation in Quantum Chromo-Dynamics \cite{cks}.

{\bf Acknowledgements} 
The authors thank H.Satz  and D.Zappala' for useful comments.

\end{document}